\documentclass[aps,prx,twocolumn,superscriptaddress,longbibliography]{revtex4-1}
\usepackage{CJK}
\usepackage{color}
\usepackage{fancyhdr} 
\usepackage{varwidth} 
\usepackage{graphicx,subfigure}  
\usepackage{bm} 
\usepackage{listings}
\usepackage{enumerate}%
\usepackage[colorlinks,citecolor=blue,linkcolor=blue]{hyperref}
\usepackage{amsmath}
\usepackage{threeparttable}
\usepackage{graphicx}
\usepackage{amssymb}
\usepackage{amsthm}
\usepackage{rotating}
\usepackage{multirow}
\usepackage{cancel} 
\def\bl#1{\textcolor{blue}{#1}}

\begin{document}
\begin{CJK*}{UTF8}{gbsn} 

\title{Ultrafast modulation of the molten metal surface tension under femtosecond laser irradiation}
\author{Chenhao Li}
\affiliation{State Key Laboratory of Precision Spectroscopy, School of Physics and Electronic Science, East China Normal University, Shanghai 200241, China}
\author{Hongtao Liang}
\affiliation{State Key Laboratory of Precision Spectroscopy, School of Physics and Electronic Science, East China Normal University, Shanghai 200241, China}
\author{Yang Yang}
\thanks{yyang@phy.ecnu.edu.cn}
\affiliation{State Key Laboratory of Precision Spectroscopy, School of Physics and Electronic Science, East China Normal University, Shanghai 200241, China}
\author{Zhiyong Yu}
\affiliation{State Key Laboratory of Precision Spectroscopy, School of Physics and Electronic Science, East China Normal University, Shanghai 200241, China}
\author{Xin Zhang}
\affiliation{State Key Laboratory of Precision Spectroscopy, School of Physics and Electronic Science, East China Normal University, Shanghai 200241, China}
\author{Xiangming Ma}
\affiliation{State Key Laboratory of Precision Spectroscopy, School of Physics and Electronic Science, East China Normal University, Shanghai 200241, China}
\author{Wenliang Lu}
\affiliation{State Key Laboratory of Precision Spectroscopy, School of Physics and Electronic Science, East China Normal University, Shanghai 200241, China}
\author{Zhenrong Sun}
\affiliation{State Key Laboratory of Precision Spectroscopy, School of Physics and Electronic Science, East China Normal University, Shanghai 200241, China}
\author{Ya Cheng}
\thanks{ycheng@phy.ecnu.edu.cn}
\affiliation{State Key Laboratory of Precision Spectroscopy, School of Physics and Electronic Science, East China Normal University, Shanghai 200241, China}

\begin{abstract}

We predict ultrafast modulation of the pure molten metal surface stress fields under the irradiation of the single femtosecond laser pulse through the two-temperature model molecular-dynamics simulations. High-resolution and precision calculations are used to resolve the ultrafast laser-induced anisotropic relaxations of the pressure components on the time-scale comparable to the intrinsic liquid density relaxation time. The magnitudes of the \bl{dynamic surface tensions} are found being modulated sharply within picoseconds after the irradiation, due to the development of the nanometer scale non-hydrostatic regime behind the exterior atomic layer of the liquid surfaces. The reported novel regulation mechanism of the liquid surface stress field and the \bl{dynamic surface tension} hints at levitating the manipulation of liquid surfaces, such as ultrafast steering the surface directional transport and patterning.

\end{abstract}
\maketitle
\end{CJK*}

The liquid surface tension plays a crucial role in advanced manufacturing\cite{Ragelle18,Khairallah16}, microfluidics\cite{Terry79}, and different chemical/biological engineering scenarios\cite{Gennes85,Huh10}. In contrast to the modern ultrafast modulation of the solid-state electronic properties\cite{Kuo17,Madeo20} or domain walls\cite{Yang20}. The manipulation of the liquid surface has long been limited on the macroscopic space and time scales\cite{Sheng14,Girot19,Liu21}. A few recent studies achieved to quantitative manipulate interfacial structure or kinetics (i.e., grain boundary migration\cite{Cash18,Wei21} or cavitation\cite{Rossello21}) via depositing precise dosing of laser/electron beam energy. However, these atomic interface regulations are relatively slow, on timescales ranging from $\mu$s to minute, leading to insufficient capability for tuning various microscopic processes on the timescale of (sub-)picosecond.

Over the last decade, femtosecond laser (fs-laser) has been widely used in advanced micro-nano processing and manufacturing\cite{Sugioka13,Malinauskas16,Saha19}.
The energy carried by the fs-laser pulse is delivered precisely to materials on a timescale much faster than conventional laser pulses, yielding either ultrafast gentle surface modification or drastic material removal through ablation. Within the strongly non-equilibrium state induced by the fs-laser irradiation, the thermal-physical properties of the warm dense matter undergo ultrafast and complicated variational fashions\cite{Zhigilei09,Wu14,Wu22}, which have not been comprehensively understood yet. Until now, few studies have focused on precisely manipulating the liquid surface via utilizing the fs-laser irradiation.

In this letter, we predict picoseconds(ps)-modulation of the molten metal surface tension under the irradiation of the single pulse fs-laser. Through employing the atomistic simulation, we calculate the high-resolution temporal evolution with sufficient statistical precision of the \bl{dynamic surface tensions} of the three molten metal surfaces irradiated with a single fs-laser pulse (and ps-laser pulse as a comparison). 
                       We use the instantaneous pressure components and the stress fields across the irradiated liquid surfaces to interpret the resulting ultrafast surface tension modulation mechanism. Ultimately, this study unveils a potential possibility of ultrafast regulating the atomic liquid surfaces, e.g., directional transport and patterning, through relatively low dose fs-laser irradiation.

The simulations of the laser irradiation of molten metal surfaces are realized by employing the two-temperature model, and molecular-dynamics hybrid method (TTM-MD)\cite{Ivanov03,Ivanov03-prb,chen15}, and performed using LAMMPS\cite{Plimpton95}. The simulation is performed for molten metal surface irradiated with a 200 fs (50 ps) laser pulse at an absorbed fluence of 16 mJ/cm$^2$. The interatomic interactions are described by the classical many-body potentials parametrized for Al\cite{Zope03}, Ti\cite{Zope03}, and Ni\cite{Pun09}, respectively. The thermodynamics properties used in modeling the electronic subsystem are chosen based on the experimental measurement and electronic structure calculations\cite{lin10}. The dimensions of the simulation domain are 75\AA$\times$75\AA$\times$1400\AA, with the periodic boundary condition applied along $x$ and $y$ axis (parallel to the irradiated surface), together with a semi-infinite boundary for mimicking pressure wave propagation and heat diffusion into the bulk liquids. The molten metal surfaces are equilibrated in the constant volume, constant temperature simulations at 0.95$T_\mathrm{m}$ up to 150 ns. Here, $T_\mathrm{m}$ represents the melting point of the studied metal. These are followed by simulations of applying laser irradiation. 500 replica TTM-MD runs are used to improve statistics. The non-equilibrium surfaces are characterized by determining of interfacial profiles, which show the instantaneous change in specific properties (e.g., density, pressure components, and stress) as functions of the distance normal $z$ to the surface plane and the temporal evolution of the \bl{dynamic surface tensions}. For additional details as to the TTM-MD model, methods of simulation setup, and analysis, see the Supplemental Material\cite{SM}.

Fig.\ref{fig1} shows the TTM-MD simulation snapshots for a molten metal surface (Al) irradiated with a single ultrashort laser pulse (pulse duration 200 fs) at an absorbed fluence of 16 mJ/cm$^2$. The laser light is absorbed by the electrons in the conduction band, introducing a thermally non-equilibrium state between the electrons and ions with different temperatures\cite{Kaganov57}. A $\sim$20 ps period following the radiation establishes the local thermal equilibrium between electrons and ions, or the two temperatures, i.e., $T_\mathrm{ion}(t)$ (black) and $T_\mathrm{e}(t)$ (red) in the Supplement Sec.III\cite{SM}, Fig.S2(a). Within this ultrashort non-equilibrium state, accompanied with the energy transfer from electrons to atomic vibrations, slight expansions (around 2 nm) of the irradiated surfaces\cite{Rethfeld17} along surface normal are seen, see Fig.\ref{fig1}. Compared with the fs-laser irradiation, both the temperature and surface expansion evolutions require nearly an order of magnitude longer, under ps-laser irradiation (pulse duration 50 ps, same absorbed fluence, Fig.S2(b)).             Here, the molten metal surfaces remain integrity throughout the irradiation simulation. Due to a relatively small fluence being chosen, the absorbed energy is smaller than threshold energies for void nucleation or spallation\cite{Upadhyay08}.

\begin{figure}[!htb]
\centering
\includegraphics[width=0.45\textwidth]{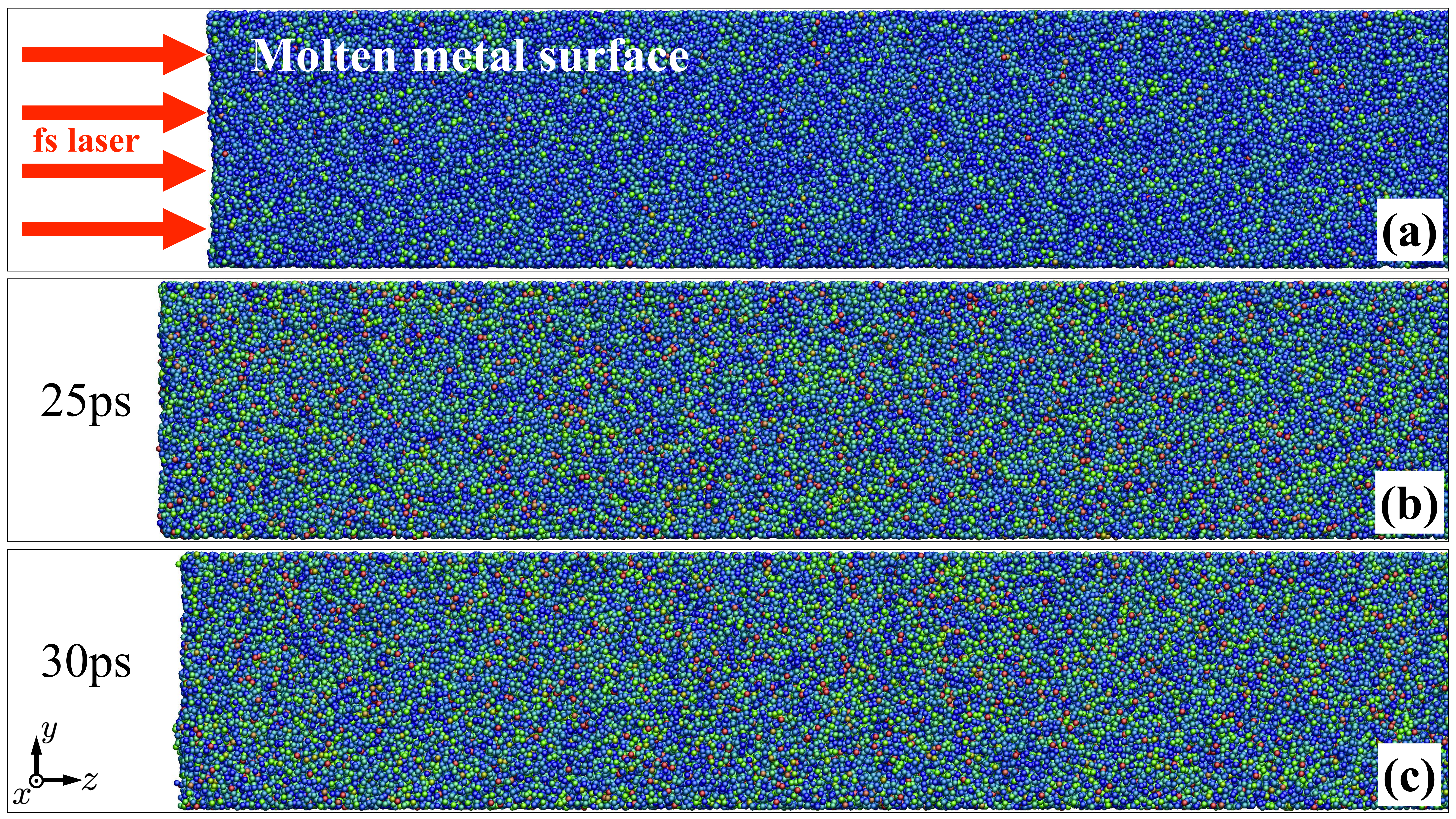}
\caption{(a)-(c) Selected snapshots from TTM-MD simulation of the molten Al surface (at 0.95$T_\mathrm{m}$, $T_\mathrm{m}$ is the melting point) irradiated with 200 fs laser pulse at an absorbed fluence of 16 mJ/cm$^2$. The laser pulse is directed along the $z$ axis, perpendicular to the molten metal surfaces from the left of plots. Particles are color-coded according to their kinetic energies, red, green and blue correspond to the high, middle and low kinetic energy, respectively.}

\label{fig1}
\end{figure}

\begin{figure}[!htb]
\centering
\includegraphics[width=0.5\textwidth]{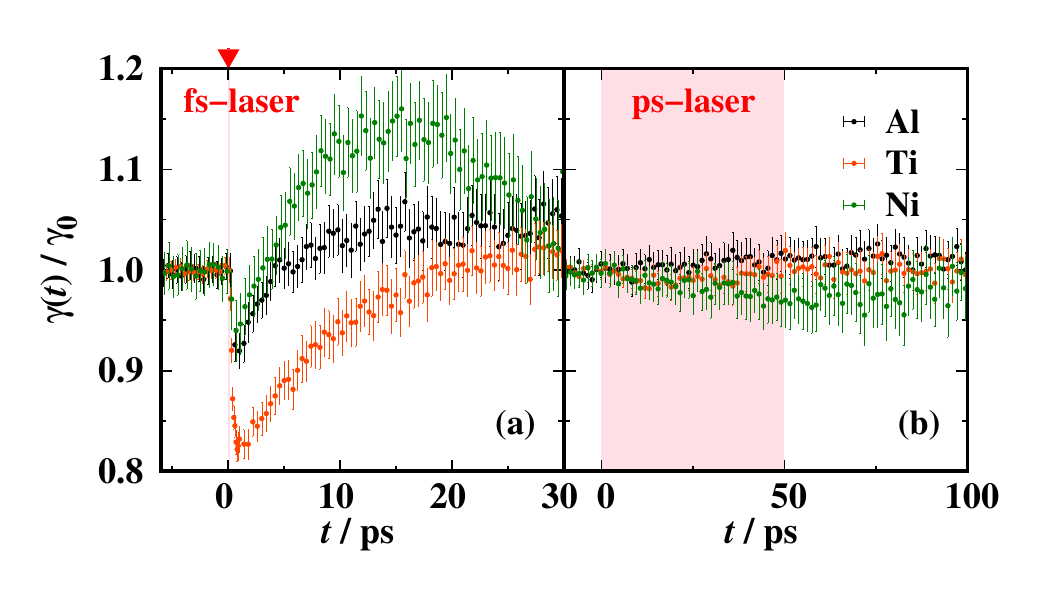}
\caption{\bl{Dynamic surface tensions} for molten Al (grey), Ti (orange), and Ni (green) irradiated by laser with two different pulse durations (denoted by pink shading areas), 200 fs (a) and 50 ps (b), both at an absorbed fluence of 16 mJ/cm$^2$. Data are normalized by their equilibrium values $\gamma_0$, Al 593(2) mN/m, Ti 1005(2) mN/m, and Ni 892(4) mN/m\cite{SM}, before laser irradiation. The error bars represent the 95\% confidence level in statistical averaging over samples from 500 independent replica simulations.}
\label{fig2}
\end{figure}

\begin{figure*}[!htb]
\centering
\includegraphics[width=0.99\textwidth]{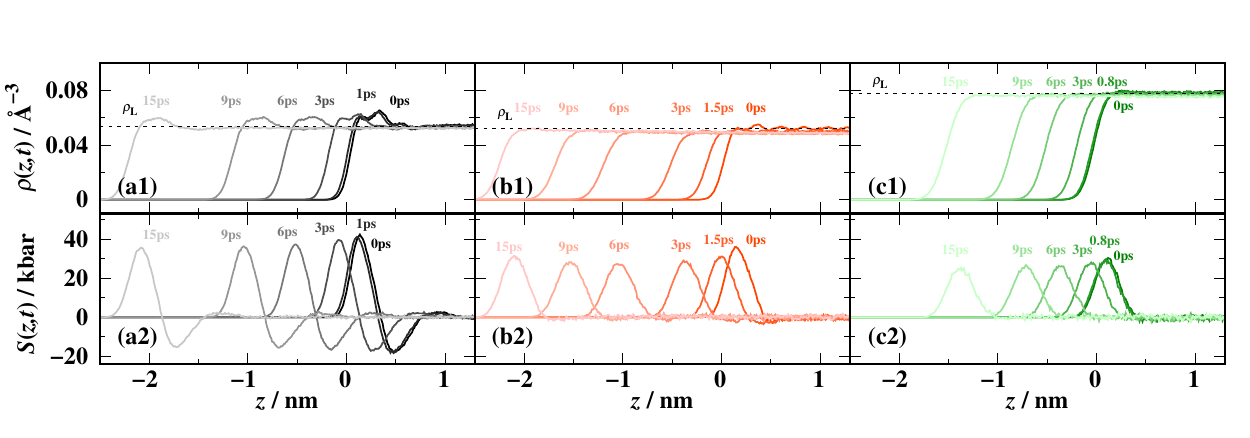}
\caption{Evolution of the density $\rho(z,t)$ and stress $S(z,t)$ profiles across the 200 fs laser pulse irradiated molten Al (a1-a2), Ti (b1-b2) and Ni (c1-c2) surfaces, at selected times. Surfaces expand along $-z$, opposite to the laser incident direction. $z=0$ corresponds to the Gibbs dividing surface position (defined with zero excess particle number density) of each equilibrium molten metal surfaces before laser irradiation.}
\label{fig3}
\end{figure*}

High-resolution and precision temporal evolutions of the \bl{dynamic surface tensions} of the irradiated molten metal surfaces irradiated by ultrafast lasers are shown in Fig.\ref{fig2}. In panel (a), distinctive changes in the \bl{dynamic surface tensions} induced by the fs-laser are observed. The normalized \bl{dynamic surface tension} $\gamma(t)/\gamma_0$ experiences a drastic drop in the first 1.0$\sim$1.5 picosecond after the laser irradiation. The transient drops for Al and Ti surfaces are followed by relatively longer periods (about 10$\sim$30 ps) relax to stable values. For Ni surface, the transient drop in $\gamma(t)/\gamma_0$ is followed by a significant elevation until a maximum excursion of 1.15$\gamma_0$ at 15 ps is reached before the relaxation stage. Note that, the variation magnitudes for the three molten metal surfaces, within the first 15 ps could reach over 20\% magnitude of $\gamma_0$, the \bl{dynamic surface tensions} can be tuned to unexpectedly lower or higher value (compared to their equilibrium values around $T_\mathrm{m}$) under the current single-shot fs-laser. \bl{Despite the heating due to laser-deposited energy, the dynamic surface tension variations shown in Fig.\ref{fig2}(a) do not comply with the known thermodynamics condition (or temperature)-dependences of the equilibrium surface tensions of the equilibrium elemental liquid-vapor interfaces\cite{Molina07,Zhou15,Wang09}, see Supplement Sec.IV\cite{SM}, Fig.S11.}

To interpret the ultrafast surface tension variation in Fig.\ref{fig2}(a), we first examine the spatial-temporal evolution of the ion density and stress profiles across the expanding surfaces for the three molten metals. In Fig.\ref{fig3}(a1)-(c1), after the fs-laser irradiation, significantly adjustment in local particle-packing are reflected from the structural variation in $\rho(z,t)$ profiles. i) Monotonically decrease in densities in the region a few nm behind the surface layer for the three metals. ii) A rapid decay of the oscillatory mode amplitude at Al and Ti surfaces indicates that the surface layering becomes weaker as the surface expands. These adjustments are akin to those of the equilibrium surfaces upon increasing the temperatures (Supplement Sec.IV\cite{SM}, Fig.S10).

Fig.\ref{fig3}(a2)-(c2) reveal the correspondingly stress profiles. Here, the stress $S$ is defined as the difference between the normal ($p_\mathrm{N}$) and transverse components ($p_\mathrm{T}$) of the pressure tensor (see Supplement Sec.II\cite{SM}). The positive stress peaks in the three metal surfaces suggest that all three surfaces are under lateral tension, whereas a part of the Al surface is under lateral compression (negative stress). Slight changes in the amplitude and width of the major positive peaks, corresponding primarily to the outermost layer particles, as time delays are noticed. For the region behind the positive peak, the variation fashion in the stress distribution is found material-dependent and more complicated. For molten Al surface, the stress regulation due to fs-laser irradiation is limited within the surface negative peak regime. In contrast, for molten Ti and Ni surfaces, the magnitudes of the stress over a broader extent ($\sim$ 20 nm behind the positive peak) experience a breathing mode, i.e., 1 ps decrease followed by a few ps increases. See more details in Supplement Sec.III\cite{SM}, Fig.S3,S6.

\begin{figure}[!htb]
\centering
\includegraphics[width=0.48\textwidth]{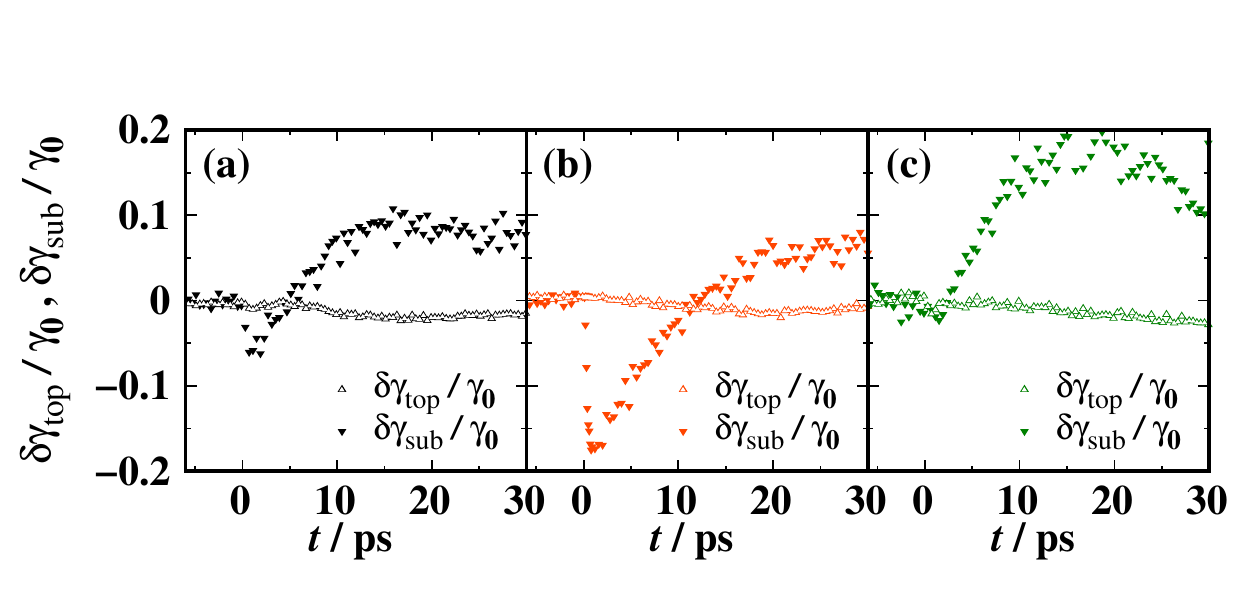}	
\caption{Temporal evolution of two contributing components of the \bl{dynamic surface tension}, for molten Al (a), Ti (b), and Ni (c) surfaces irradiated with the 200 fs laser pulse. The data of the two contributing components are represented with the differences ($\delta\gamma_\mathrm{top}(t)=\gamma_\mathrm{top}(t)-\gamma_\mathrm{0top}$, $\delta\gamma_\mathrm{sub}(t)=\gamma_\mathrm{sub}(t)-\gamma_\mathrm{0sub}$) between the temporal instantaneous values and the equilibrium values, normalized by $\gamma_0$. $\gamma_\mathrm{0top}$ for three equilibrium molten surfaces are Al 985(1) mN/m, Ti 1050(1) mN/m, and Ni 896(1) mN/m, respectively. Correspondingly, $\gamma_\mathrm{0sub}$, Al 392(1) mN/m, Ti -45(1) mN/m, and Ni -4(3) mN/m.}
\label{fig4}
\end{figure}

To gain further quantitative insight, we separately count the local contributions through dividing the Kirkwood-Buff equation (mechanical definition of the surface tension)\cite{Kirkwood49} into two components, $\gamma(t) = \gamma_\mathrm{top}(t)+\gamma_\mathrm{sub}(t) = \int_{z_\mathrm{lo}}^{z_1(t)} S(z,t) \mathrm{d}z + \int_{z_1(t)}^{z_\mathrm{hi}} S(z,t) \mathrm{d}z$. The contribution of the outermost atomic layer (positive stress peak)\cite{Yuan22}, $\gamma_\mathrm{top}$, is distinguished from the contribution of the rest region of the surface, $\gamma_\mathrm{sub}$. $z_1(t)$ is the position where positive peak ends.  For all three metal systems, it is found that $\gamma_\mathrm{top}(t)$ forms the fundamental magnitude basis of $\gamma(t)$, whereas $\gamma_\mathrm{sub}(t)$ dominates the variations, see in Fig.\ref{fig4}. \bl{This mechanism completely departs from that of the traditional thermodynamics condition (or equilibrium temperature) regulation on the surface tension, in which the surface becomes broader and diffuse as the liquid-vapor critical temperature is approached, $\gamma_\mathrm{top}(T)$ decreases monotonically with temperature $T$ along with the diminishing surface layering, see Supplement Sec.IV\cite{SM}, Fig.S10,S11.} Therefore, a novel \bl{dynamic surface tension} manipulation mechanism can be inferred. The local surface particle packing is regulated within the extreme non-equilibrium stage after the fs-laser irradiation. The local stress field is modified with bias, yielding an ultrashort modulation of the dynamic surface tension and unexpectedly significant magnitude, which may be inaccessible for equilibrium liquid surfaces.

Interestingly, in contrast to the fs-laser induced variations in Fig.\ref{fig2}(a), under irradiation of single-shot ps-laser (pulse duration 50 ps) at the same absorbed fluence, within statistical uncertainties, $\gamma(t)/\gamma_0$ data of the three metal systems remain almost unchanged throughout the simulation. see Fig.\ref{fig2}(b). By examining the density, stress profiles, $\gamma_\mathrm{top}(t)$ and $\gamma_\mathrm{sub}(t)$ for these ps-laser irradiated surfaces, see Fig.S4,S5,S7,S8. Evident regulation of the surface stress distributions, as the fs-laser cases, are not found. The thermal expansions of the ps-laser irradiated surfaces accompanied by density variations occur. However, they take more than 100 ps to reach the same magnitude of expansion extent caused by fs-laser in about ten ps. Therefore, under the same absorbed fluence in the current study, the surface stress field being regulated only by fs-laser is seemingly related to the density variation rate rather than the density itself.

To support the above speculation, a characteristic collective dynamics time-scale, i.e., the so-called liquid density relaxation times $\tau$, defined as the inverse half-width of the dynamic structure factor, is determined independently from equilibrium bulk liquid at $T_\mathrm{m}$, see Supplement Sec.IV\cite{SM}, Fig.S11. The obtained $\tau$ for Al, Ti, and Ni are 0.57(6) ps, 0.56(4) ps, and 0.53(4) ps, respectively. The ultrafast surface tension regulation time, due to fs-laser pulse, is comparable to $\tau$, implying a higher probability that the natural particle-packing relaxation path is altered by the ultrafast energy deposition anisotropically. By contrast, the ps-laser-induced regulation in particle density takes about two orders of magnitude longer than $\tau$ under ps-laser irradiation, which is too long to rig the intrinsic particle packing rearrangement process. Hence, the original surface mechanical balance is not altered during thermal expansion, which is identical to the liquid surfaces' equilibrium-state heating process.

\begin{figure}[!htb]
\centering
\includegraphics[width=0.48\textwidth]{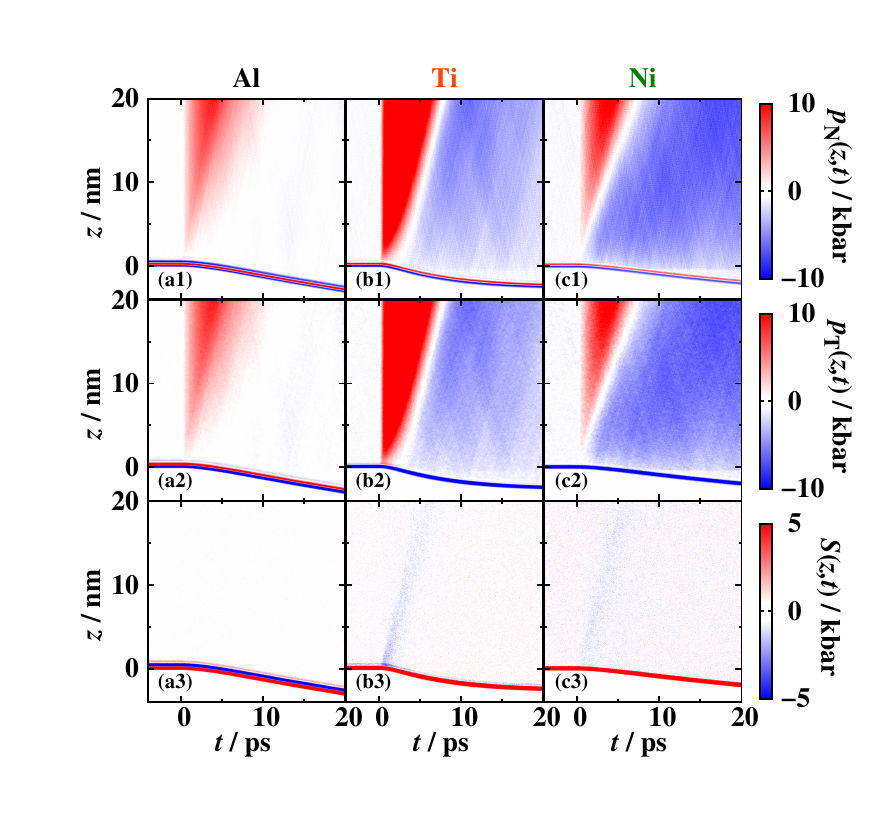}
\caption{Contour plots of fine-grained profiles of the normal ($p_\mathrm{N} (z,t)$) and transverse components ($p_\mathrm{T} (z,t)$) of the pressure tensor, and the stress ($S (z,t)=p_\mathrm{N} (z,t)-p_\mathrm{T} (z,t)$) across the molten Al, Ti, and Ni surfaces irradiated with the 200 fs laser pulse. The laser pulse is directed along $z$ axis, perpendicular to the molten metal surfaces from the bottom of the contour plots.}
\label{fig5}
\end{figure}

Fig.\ref{fig5}(a1)-(c2) show contour plots of the temporal-spatial evolution of the transverse and normal pressure component profiles. Right after the fs-laser pulse, positive pressure immediately builds up, in both the $p_\mathrm{N} (z,t)$ and $p_\mathrm{T} (z,t)$, due to the fast heating and the insufficient time for ions to expand. At a time delay of $\sim$ 0.5 ps after laser irradiation, the pressure starts to decay, accompanying the initiation of the surface expansion, a dynamic transition region (propagating inwards of bulk liquid) in which positive pressure decays to zero or even negative. It is known that the ultrafast laser-induced inertial confinement of high internal pressures, along with the structural phase transition and the thermal expansion of warm dense matter, have been reported extensively\cite{shugaev18,Ivanov03,Rethfeld17,Rossello21,Wu22}. However, to the best of the authors' knowledge, few studies have been carried out to explore the anisotropic evolution among pressure components.

For the Ti and Ni systems, finite differences between the two pressure components can be identified and be also visualized in the contour plots of the temporal-spatial stress profile, see $S(z,t)$ in Fig.\ref{fig5}(b3),(c3). The development of the finite stress in the 10$\sim$20 nm region behind the outermost surface layer \bl{indicates the broadening of the local non-hydrostatic region on the basis of the original few interfacial layers with non-zero stresses.} It explains the $\gamma_\mathrm{sub}(t)$ variation in Fig.\ref{fig4}(b),(c). The negative (or positive) stress corresponds to the fact that the sub-surface liquids are in the state of relative lateral tension (or compression), even if thet two pressure components are both positive or both negative. These anisotropic variations between two pressure components support the above inference of the particle-packing relaxation being disturbed anisotropically by the fs-laser irradiation. In addition, a visible difference among materials being noticed, i.e., in molten Al surface, Fig.\ref{fig4}(a), the anisotropic pressure variation in response to the ultrafast thermal expansion and pressure release, are confined within a thin-film region corresponding to the negative stress peak - 0.5$\sim$1.0 nm next to the outermost atomic surface layer. It seems that this thin-film region at the molten Al surface has a higher priority in absorbing the shock wave induced by the ultrafast laser pulse, and preventing the adjacent liquid phase particles from undertaking finite stresses, see in Fig.\ref{fig5}(a3). We should note that, in order to make a conclusion about such wave attenuation mechanism, one needs to carry out a standalone future study, i.e., the study performed with manually generated acoustic waves with well-controlled properties rather than ones produced by laser irradiation.

By contrast, under the irradiation of the ps-laser, visible differences between the two pressure components in the molten metal surfaces can not be identified, see Supplement Sec.III\cite{SM}, Fig.S9. Therefore, it is evident that laser pulse with a relatively long duration is incapable of disrupting the alteration of the particle packing, thus modulating the liquid \bl{dynamic surface tension} on the sub-ps time-scale of the liquid density relaxation.

In summary, by employing the TTM-MD simulations and high-precision statistical calculations, we predict that the \bl{dynamic surface tensions} of the pure molten metals being modulated sharply  within a few picoseconds under the (relatively low dose) irradiation of single-pulse of fs-laser. Within this ultrafast modulation, liquid phase particles behind the exterior atomic layer receive the deposited laser energy on the time-scale comparable to the time-scale of their intrinsic collective dynamics (density relaxation time). The surface expansion occurs concurrently, leading to a biased development among different pressure components, and finally result in the ultrafast variation in the surface stress distribution and/or development of the \bl{local non-hydrostatic region} within a lengthscale of a few tens of nm behind the liquid surfaces. However, our irradiation simulations suggest that ps-laser pulse at the same absorbed fluence could not tune the local surface mechanical properties under a similar microscopic mechanism because the ps-laser pulse duration is too long to impact the particle packing relaxation kinetics.

\bl{For an equilibrium liquid-vapor interface system, the bulk liquid phase and bulk vapor phase outside the interface area naturally satisfy the hydrostatic pressure condition with zero stresses. Meanwhile, there are 2-3 layers of particles in the liquid-vapor interfacial transition region, where the stresses are not equal to 0. As shown in the stress profiles in Fig.\ref{fig3} ($t=0$ ps, near $z=0$ nm), particles in this local region are intrinsically non-hydrostatic, which relates to the origin of the surface tension, in other words, even for the liquid-vapor interface that satisfies the hydrostatic equilibrium, the intrinsic local non-hydrostatic region still exists within.\cite{Rowlinson82}} From a certain point of view, the pure molten metal surfaces could be thickened in picoseconds by the fs-laser irradiations, e.g., broadening the \bl{local non-hydrostatic region} for molten Ti and Ni surfaces, meanwhile, the raw mechanical scenario within the outermost layer remains nearly unaltered. \bl{The currently reported manipulation mechanism completely departs from traditional thermodynamics condition (or equilibrium temperature) regulation on surface tension. It would usher in a new level of ultrafast engineering of the liquid-vapor interfacial stress field and dynamic surface tension and potentially levitate the ability to steer the directional transport and patterning of the molten metal surfaces.}

The authors acknowledge the National Key R\&D Program of China (2019YFA0705000), the Chinese National Science Foundation (Grant No. 11874147, 11933005, 12134001), the Science and Technology Commission of Shanghai Municipality (21DZ1101500), Shanghai Municipal Science and Technology Major Project (2019SHZDZX01), Natural Science Foundation of Chongqing, China (Grant No. cstc2021jcyj-msxmX1144), and the State Key Laboratory of Solidification Processing in NWPU (Grant No. SKLSP202105).

\bibliography{ref}

\end{document}